\documentclass[pre,twocolumn,showpacs,amsmath,eqsecnum,amssymb]{revtex4}
\newcommand{\be}{\begin{equation}} 
\newcommand{\ee}{\end{equation}} 
\newcommand{\bea}{\begin{eqnarray}} 
\newcommand{\eea}{\end{eqnarray}}
\usepackage{graphicx}
\usepackage{bm}

\begin{document}
\title{\large{Langevin Simulation of the Chirally Decomposed Sine-Gordon Model}}
\author{L. Moriconi and M. Moriconi}
\affiliation{Instituto de F\'\i sica, Universidade Federal do Rio de Janeiro, \\
C.P. 68528, 21945-970, Rio de Janeiro, RJ, Brazil}

\begin{abstract}
A large class of quantum and statistical field theoretical models, encompassing relevant condensed matter and non-abelian gauge systems, are defined in terms of complex actions. As the ordinary Monte-Carlo methods are useless in dealing with these models, alternative computational strategies have been proposed along the years. The Langevin technique, in particular, is known to be frequently plagued with difficulties such as strong numerical instabilities or subtle ergodic behavior. Regarding the chirally decomposed version of the sine-Gordon model as a prototypical case for the failure of the Langevin approach, we devise a truncation prescription in the stochastic differential equations which yields numerical stability and is assumed not to spoil the Berezinskii-Kosterlitz-Thouless transition. This conjecture is supported by a finite size scaling analysis, whereby a massive phase ending at a line of critical points is clearly observed for the truncated stochastic model.
\end{abstract}

\pacs{05.50.+q, 05.10.Gg, 11.10.Wx}

\maketitle

\section{Introduction} 
Several field theory models devoted to concrete condensed matter and high energy applications are given in terms of complex-valued bosonic (``c-number") actions. In the condensed matter context, for instance, they are the focus of much attention in low-dimensional systems related to Luttinger liquids and quantum Hall phenomena \cite{kane,pryadko,naud,kim}. On the high-energy front, on the other hand, a fundamental open issue refers to the phase diagram of SU(3)-QCD with non-zero fermion density, where the gauge action notoriously acquires an imaginary part, due to the non-vanishing quark chemical potential
\cite{karsh,barbour,ambjorn,bilic,kogut,delia,muroya}.

Numerical non-perturbative efforts to study all of the above problems are considerably restricted by the fact that the Monte-Carlo method cannot be straightforwardly applied, since the weights in the path-integration measures are not positive-definite quantities anymore. To overcome this difficulty, a number of alternative numerical methods have been suggested, from time to time, such as complex Langevin simulations \cite{parisi,klauder}, analytical expansions \cite{roberge}, kernel mappings \cite{okamoto}, microcanonical averaging prodecures \cite{baaquie}, path-integral factorization schemes \cite{ambjorn2}, etc. To date, there are not yet well-established results on how general and efficient these tools are. Among them, the Langevin approach -- the central subject of discussion in this paper -- has as attractive features its great simplicity and direct numerical implementation. However, as it has been learned from purely empirical investigations, the stochastic evolution provided by the Langevin equations is affected, in many important cases, by bad numerical convergence and the existence of blow-up trajectories \cite{kogut,gausterer}. Also, some special care may be needed when writing the Langevin equations for actions which have cuts in the complex plane, in order to assure good ergodic properties of the phase space flow \cite{ambjorn,fujimura}.

It is not hard to understand in an essential way the origin of eventual instabilities in complex Langevin equations. Since the field variables are promoted to complex numbers in the stochastic differential equations, the configurational phase space is enlarged, and unstable manifolds appear along the imaginary direction, containing the classical vacuum solutions. The time evolution of a phase space point (that is, a field configuration) will depart, in brownian-like motion, from the deterministic trajectory that would be followed in the absence of the stochastic noise. Thus, as soon as the phase space point happens to visit a region of ``intense stream" in the neighborhood of an unstable manifold, the numerical convergence of the Langevin equation is under risk. 

While in polynomial models like the $\lambda \phi^4$ with complex coupling constants the Langevin simulation is mostly successful in the cases of physical interest, difficulties arise in other well-known models where the self-interaction potential is non-polynomial and invariant under compact Lie groups. The Langevin equations will contain, due to the extension of fields to the complex plane, ``explosive" hyperbolic functions, leading, in general, to unstable numerical simulations \cite{gausterer}. That is precisely the situation found in many of the bosonized condensed matter systems or in the finite Baryon density QCD realm. 

In order to investigate the problem of numerical Langevin stability, we have regarded the exactly solvable euclidean sine-Gordon (SG) model as an ideal testing ground prototype, aiming, in this simpler context, to antecipate applications for a large class of systems. As briefly reviewed in Sec. II, the SG partition function is alternatively written in terms of interacting left and right chiral bosons, governed by a complex-valued action. In Sec. III, we introduce the stochastic quantization procedure for the chirally decomposed SG model, which is observed to produce numerical blow-up solutions. It turns out that the instabilities are not fixed by a reasonable time step resizing, or even with the help of more precise evolution algorithms (as a fourth-order Runge-Kutta scheme). We propose, then, a pragmatical solution of the numerical convergence problem, through a straightforward ``truncation prescription" applied to the stochastic equations. The basic idea is to modify the Langevin equations without spoiling its thermodynamical phases. In fact, as a strong piece of evidence supporting our method, the Berezinskii-Kosterlitz-Thouless (BKT) phase transition \cite{ber,ko-th} of the original SG model is verified by means of a finite size scaling study. Finally, in Sec. IV, we comment on our findings and discuss directions of further research.

\section{The Chirally Decomposed SG Model}

As it is well-known, the usual euclidean SG correlation functions may be obtained, in path-integral language, from the zero temperature partition function
\cite{zinn-justin},
\be
Z = \int D \phi \exp  \{- \int d^2x  [ \frac{1}{2} (\partial \phi)^2
- \frac{g}{\beta} \cos(\beta \phi)  ]  \} \ . \ \label{sg}
\ee
It is worth summarizing here some important facts on the above model \cite{gogolin}. In the $g \rightarrow 0$ limit, the SG spectrum is massless for $\beta^2 \geq 8 \pi \equiv \beta_c^2$, defining a line of critical points, and massive for $\beta^2 < \beta_c$. At $\beta = \beta_c$ the SG model has a BKT phase transition. The same behavior holds for finite $g$, with $\beta_c=\beta_c(g)$ ($\beta_c$ grows with $g$). The SG model may be regarded as the bosonized version of the massless Thirring model perturbed by a Gross-Neveu interaction, defined in terms of a Dirac fermion field $\psi$ \cite{fradkin}. Its phases can be distinguished by the mass order parameter $m \equiv \langle \bar \psi \psi \rangle$, which is proportional to 
$ \Phi \equiv \langle \cos( \beta \phi/2) \rangle$. For $\beta \rightarrow \beta_c$ from below, the exact solution of the SG model yields
$
\langle \bar \psi \psi \rangle \sim \exp[c/(\beta-\beta_c)] \ . \ \label{mass_thi}
$
For $\beta > \beta_c$ the SG infrared fixed point is just the gaussian model and the order parameter vanishes in the thermodynamical limit. At $\beta=\beta_c$ the order parameter has scaling dimension $\Delta (\Phi)=-1/2$, that is, for a finite size system with linear dimension $L$,
it follows that $\langle \Phi \rangle \sim L^{-1/2}$.

The chirally decomposed version of the SG model (CDSG) can be readly derived in a few elementary steps. Introducing the conjugate momentum $\pi=\pi(x_0,x_1)$, we may rewrite (\ref{sg}) as
\bea
Z &=& \int D \phi D \pi \exp  \{- \int d^2x  [\frac{\pi}{2} + i \pi \partial_0 \phi \nonumber \\
&+& \frac{1}{2} (\partial \phi)^2
- \frac{g}{\beta} \cos(\beta \phi)  ]  \} \ . \ \label{sg_b}
\eea
Define now the right and left chiral fields,
\bea
&&\phi_R = \frac{1}{2} [ \phi + \int_0^{x_1} d \xi \pi(x_0,\xi) ] \ , \ \nonumber \\
&&\phi_L = \frac{1}{2} [ \phi - \int_0^{x_1} d \xi \pi(x_0,\xi) ] \ . \ \label{chi-fields}
\eea
It is straightforward to invert (\ref{chi-fields}). We find
\bea
&&\phi = \phi_R + \phi_L \ , \ \nonumber \\
&&\pi = \partial_1 \phi_R - \partial_1 \phi_L \ . \
\eea
As the jacobian of the mapping $(\phi,\pi) \rightarrow (\phi_R,\phi_L)$ does not depend on the fields, the partition function 
(\ref{sg_b}) may be expressed, by means of a direct substitution, as
\be
Z = \int D \phi_R D \phi_L 
\exp (- S ) \ , \
\ee
where
\bea
S &=& \int d^2x [ \partial_1 \phi_R (\partial_1 + i \partial_0) \phi_R \nonumber \\
&+& \partial_1 \phi_L (\partial_1 - i \partial_0) \phi_L
- \frac{g}{\beta} \cos(\beta (\phi_R+\phi_L))]   \label{cdsg-s} 
\eea
is, thus, defined as the action of the CDSG model. It is clear that correlation functions of the SG model can be computed, in principle, from the above partition function, through the substitution $\phi \rightarrow \phi_R + \phi_L$. Observe, however, that the CDSG action is complex, and the numerical chiral correlation functions cannot be obtained with the help of ordinary Monte-Carlo techniques. As an alternative numerical method, we will investigate the CDSG model, in the next section, from the point of view of Langevin equations, based on the formalism of stochastic quantization.

\section{Stochastic Simulations}
The Parisi-Wu stochastic quantization method \cite{parisi-wu}, states, under very general hypothesis, that euclidean field theories can be modelled through Langevin (= stochastic) differential equations. Actually, the CDSG model is equivalently given by the following set of Langevin equations:
\bea
&&\partial_\tau \phi_R = - \frac{\delta S}{\delta \phi_R} + \eta_R  \ , \ \nonumber \\
&&\partial_\tau \phi_L = - \frac{\delta S}{\delta \phi_L} + \eta_L  \ , \ \label{lang-eqs}
\eea
where $\tau$ is the auxiliary ``stochastic time". In the asymptotic $\tau \rightarrow \infty$ limit, we expect that correlation functions involving the chiral fields, $\phi_R(\tau,x_0,x_1)$ and $\phi_L(\tau,x_0,x_1)$, taken at equal stochastic times, will converge to the CDSG ones. Both $\eta_R$ and 
$\eta_L$ are gaussian random fields, satisfying to
\bea
&&\langle \eta_R \rangle = \langle \eta_L \rangle = 0 \ , \ \nonumber \\
&&\langle \eta_R(\tau,x) \eta_L(\tau',x') \rangle = 0 \ , \ \nonumber \\
&& \langle \eta_R(\tau,x) \eta_R(\tau',x') \rangle = 2 \delta(\tau - \tau')\delta^2(x-x') \ , \ \nonumber \\
&&\langle \eta_L(\tau,x) \eta_L(\tau',x') \rangle = 2 \delta(\tau - \tau')\delta^2(x-x') \ . \ 
\eea
Above, $x$ and $x'$ denote points in the physical two-dimensional space. Substituting the action (\ref{cdsg-s}) into (\ref{lang-eqs}) we get
\bea
&&\partial_\tau \phi_R  =
2\partial_1(\partial_1 + i \partial_0) \phi_R - g \sin(\beta \phi) + \eta_R  \ , \ \nonumber \\
&&\partial_\tau \phi_L=
2\partial_1(\partial_1 - i \partial_0) \phi_L - g \sin(\beta \phi) + \eta_L \ , \ \label{lgeqs}
\eea
where $\phi = \phi_R + \phi_L$. Since these equations are complex, it is convenient, for simulation purposes, to split their real and imaginary components. Define $\phi_R= \phi_R^{(1)}+i \phi_R^{(2)}$ and $\phi_L = \phi_L^{(1)}+i\phi_L^{(2)}$. The resulting equations are, therefore,
\bea
\partial_\tau \phi_R^{(1)}  &=&
2\partial_1^2  \phi_R^{(1)} - 2 \partial_1 \partial_0 \phi_R^{(2)} \nonumber \\
&-& g \cosh(\beta \phi^{(2)}) \sin(\beta \phi^{(1)}) + \eta_R  \ , \ \nonumber \\
\partial_\tau \phi_R^{(2)} &=&
2\partial_1^2 \phi_R^{(2)} - 2 \partial_1 \partial_0 \phi_R^{(1)} \nonumber \\
&-& g \sinh(\beta \phi^{(2)}) \cos(\beta \phi^{(1)}) \ , \  \nonumber \\
\partial_\tau \phi_L^{(1)}  &=&
2\partial_1^2  \phi_L^{(1)} + 2 \partial_1 \partial_0 \phi_L^{(2)} \nonumber \\
&-& g \cosh(\beta \phi^{(2)}) \sin(\beta \phi^{(1)}) + \eta_L  \ , \ \nonumber \\
\partial_\tau \phi_L^{(2)} &=&
2\partial_1^2 \phi_L^{(2)} + 2 \partial_1 \partial_0 \phi_L^{(1)} \nonumber \\
&-& g \sinh(\beta \phi^{(2)}) \cos(\beta \phi^{(1)}) \ . \   \label{lgeqsb}
\eea

It is a hopeless task to solve the above coupled differential equations via direct numerical simulations. In general, for $\beta$ close to $\beta_c$ (but not necessarily within the critical region), the numerical convergence is ruined by the existence of blow-up configurations. That kind of phenomenon is akin to the one found in simulations of lattice QCD, when the quark chemical potential is set around the threshold for fermion production \cite{kogut}. After several atempts, we concluded it is unlikely that the numerical evolution can be stabilized by simple time step reductions or the use of improved higher order Runge-Kutta schemes. However, from a purely mathematical perspective, the source of instability in the Langevin simulation of the CDSG model is clear: it is related to the hyperbolic functions defined in equations (\ref{lgeqsb}). Their eventual fast growing, likely to happen close to phase transition points where field fluctuations are more intense, cannot be tamed by any finite precision numerical algorithm.

Relying upon the idea of universality classes, we propose a way out of the instability problem.  One may argue that as far as the correlation length is much larger than the lattice spacing, the main interest is not the precise location, in the phase diagram, of the phases of the SG model. Instead, it is important to retain, in the continuum limit, the ``topology" of the phase diagram and the correct singularity order of the transitions between different phases. In loose words, we suggest to change the model, but not the large scale physics. We have found, in fact, that it is possible to deform the above set of equations, rendering them numerically stable, while preserving the BKT transition they are assumed to imply. 

To understand how the Langevin equations are to be modified, it is instructive to consider the following ordinary differential equation, directly inspired from the form of (\ref{lgeqs}):
\be
\dot \psi = - \sin \psi \ . \ \label{psi-eq}
\ee
Extending the field $\psi$ to the complex plane, by means of $\psi = \psi_1 + i \psi_2$, equation (\ref{psi-eq}) becomes
\bea
\dot \psi_1 &=& -\sin \psi_1 \cosh \psi_2 \ , \ \nonumber \\
\dot \psi_2 &=& - \cos \psi_1 \sinh \psi_2 \ . \ \label{psi-eqsb}
\eea
The integral curves for this dynamical system, that is the vector field $(\dot \psi_1, \dot \psi_2)$, are indicated in the top of Fig. 1. Due to the large ``speeds" far from the $\psi_2=0$ axis, numerical solutions exhibit bad uniform convergence properties: for any choice of time step, a point close enough to one of the unstable manifolds would evolve towards a region of difficult numerical control.
\vspace{-0.5cm}

\begin{figure}[tbph]
\includegraphics[width=10cm, height=13cm]{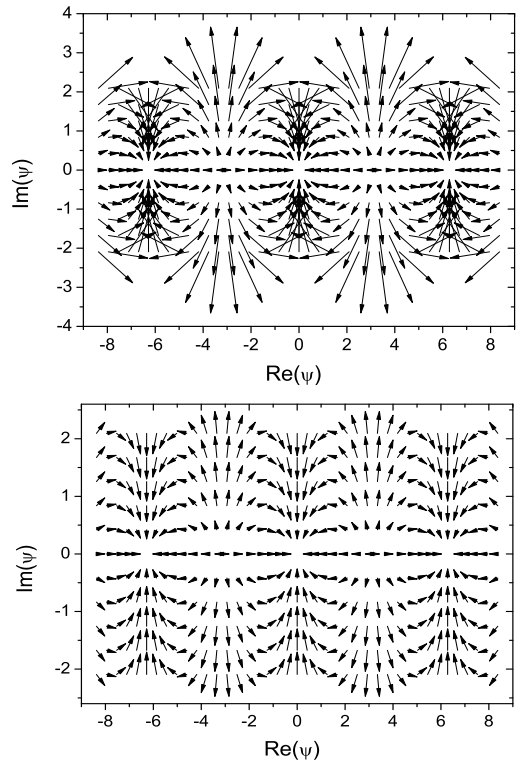}
\caption{The vector fields for the dimensionally-reduced dynamical systems (\ref{psi-eqsb}) and (\ref{psi-eqsc}).}
\label{fig1}
\end{figure}
A simple modification of (\ref{psi-eqsb}) leads to a dynamical system with uniform convergence properties, and the same asymptotic solutions. To get the alternative set of differential equations, we perform, in (\ref{psi-eq}), the substitution:
\bea
&&\sin \psi = \sin(\psi_1+i \psi_2) \nonumber \\
&&= \cosh \psi_2 \sin \psi_1 +i \sinh \psi_2 \cos \psi_1  \nonumber \\
&&= \cosh \psi_2 ( \sin \psi_1 +i \tanh \psi_2 \cos \psi_1 ) \nonumber \\
&&\rightarrow \sin \psi_1 + i \tanh \psi_2 \cos \psi_1 \ . \ \label{trunc}
\eea
Now, instead of (\ref{psi-eqsb}), we will have
\bea
\dot \psi_1 &=& -\sin \psi_1 \ , \ \nonumber \\
\dot \psi_2 &=& - \cos \psi_1 \tanh \psi_2 \ . \ \label{psi-eqsc}
\eea
The integral curves for the above model are well-behaved, as it may be seen from the picture shown at the bottom of Fig. 1. Indeed, the vector field is bounded, for $||(\dot \psi_1, \dot \psi_2)||^2 = \sin^2(\psi_1)+ \cos^2(\psi_1) \tanh^2(\psi_2) < 1$ and numerical convergence is never a problem in this framework. It is worth noting that the truncation prescription (\ref{trunc}) keeps the original structure of the stable and unstable manifolds, as well as the positions of the fixed points in the $(\psi_1,\psi_2)$ plane. Its role is basically to soften the vector field for the purpose of having stable numerical simulations.

The same stabilization trick may be applied without any further complication to the CDSG Langevin equations (\ref{lgeqs}). We get
\bea
\partial_\tau \phi_R^{(1)}  &=&
2\partial_1^2  \phi_R^{(1)} - 2 \partial_1 \partial_0 \phi_R^{(2)} \nonumber \\
&-& g \sin(\beta \phi^{(1)}) + \eta_R  \ , \ \nonumber \\
\partial_\tau \phi_R^{(2)} &=&
2\partial_1^2 \phi_R^{(2)} - 2 \partial_1 \partial_0 \phi_R^{(1)} \nonumber \\
&-& g \tanh(\beta \phi^{(2)}) \cos(\beta \phi^{(1)}) \ , \  \nonumber \\
\partial_\tau \phi_L^{(1)}  &=&
2\partial_1^2  \phi_L^{(1)} + 2 \partial_1 \partial_0 \phi_L^{(2)} \nonumber \\
&-& g \sin(\beta \phi^{(1)}) + \eta_L  \ , \ \nonumber \\
\partial_\tau \phi_L^{(2)} &=&
2\partial_1^2 \phi_L^{(2)} + 2 \partial_1 \partial_0 \phi_L^{(1)} \nonumber \\
&-& g \tanh(\beta \phi^{(2)}) \cos(\beta \phi^{(1)}) \ . \ \label{lgeqsc}
\eea

Numerical simulations were successfuly carried out for the above set of Langevin equations, using an elementary Euler evolution algorithm, for lattices with sizes $5 \times 5$, $10 \times 10$, $15 \times 15$, and $20 \times 20$. Periodic boundary conditions were imposed on the field configurations.
\vspace{-0.3cm}

\begin{figure}[tbph]
\includegraphics[width=9.5cm, height=8.5cm]{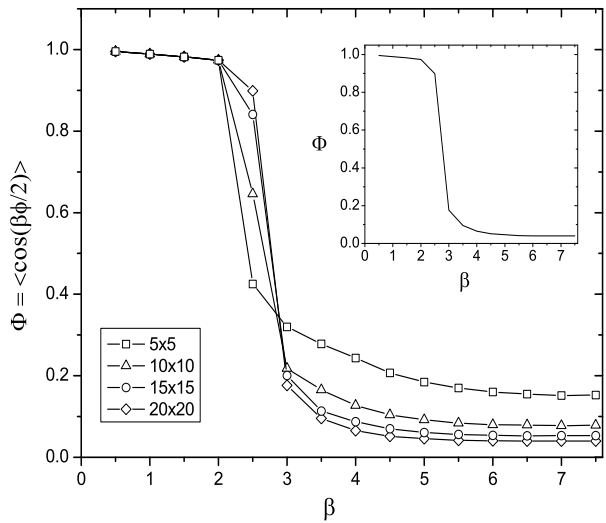}
\caption{The mass order parameter $\Phi$ is plotted as a function of $\beta$ for different lattice sizes. The inset shows the data for the lattice
with size $20 \times 20$.}
\label{fig2}
\end{figure}

For each lattice size and a given value of $\beta$, an ensemble of 6 simulations was considered, related to different seeds for the pseudorandom number generator. Each simulation consisted of $10^7$ iterations per lattice site. The SG coupling constant was set to $g=10.0$. The time step was taken to be $\epsilon = 2 \times 10^{-3}$. Averages were computed over a set of approximately decorrelated (in stochastic time) $2000$ configurations per simulation.

In Fig. 2, the mass order parameter $\Phi$ is plotted for $\beta=0.5, 1.0, \ldots , 7.5$. In more detail, we have evaluated, for a lattice of linear size $L$,
\be
\Phi = \frac{1}{6 \times 2000 \times L^2} \sum_{k=1}^{2000} \sum_{l=1}^6| \sum_{i,j=1}^L \cos(\frac{\beta \phi_k^l(i,j)}{2}) |  \ , \
\ee
where $(k,l)$ labels a given complex configuration of $\phi=\phi(i,j)$ in the statistical sample. There is clearly a phase transition taking place around $\beta_c =2.5$. The inset of Fig. 2 gives the best profile for the transition among our simulations, for the lattice with size $20 \times 20$. 
\vspace{-0.3cm}
\begin{figure}[tbph]
\includegraphics[width=9.5cm, height=8.5cm]{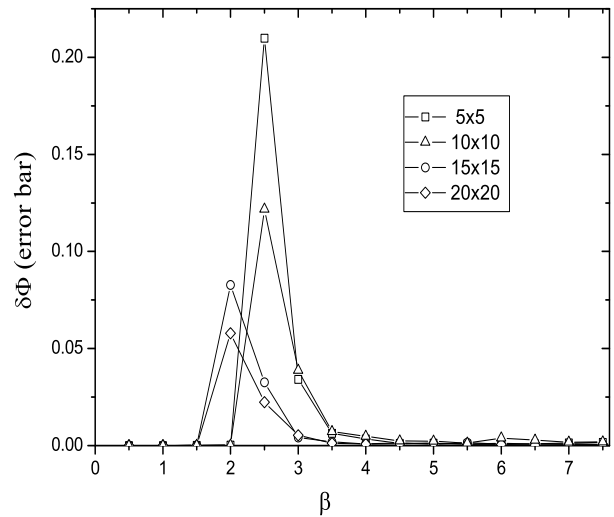}
\caption{The standard deviation analysis for the set of averaged evaluations of the mass order parameter $\Phi$.}
\label{fig3}
\end{figure}

In addition, we have carried out an estimate of the error bars related to the curves shown in Fig. 2. They are defined as
\be
\delta \Phi = \sqrt{ \frac{1}{6} \sum_{l=1}^6(\Phi_l)^2 - \Phi^2} \ , \
\ee
where
\be
\Phi_l = \frac{1}{2000 \times L^2} \sum_{k=1}^{2000}| \sum_{i,j=1}^L \cos(\frac{\beta \phi_k^l(i,j)}{2}) |  \ . \
\ee
The results are shown in Fig. 3. As expected, fluctuations are larger around the phase transition, and get smaller as the lattice size grows. For the lattice with size $20 \times 20$, errors are just of the order of a few percent.

Different lattice sizes were studied in order to establish a finite size scaling analysis of the transition, reported in Fig. 4. Each one of the five straight lines draw in that picture corresponds to a fixed value of $\beta$ ($\beta=3.5, 4.5, 5.5, 6.5, 7.5$), indicating that the scaling relation
\be
\Phi \sim L^{\Delta} \label{size-effec}
\ee
holds at the right side of the transition. According to the ideas of finite size scaling analysis \cite{zinn-justin}, equation (\ref{size-effec}) provides strong evidence for the fact that the whole region $\beta > 3.5$ is a massless phase of the model under investigation. No mass scale is involved; the only relevant scale turns to be the lattice linear size. Incidentally, as $\Delta \simeq -1$ in the extended critical region, $\Phi$ plays indeed the role of a {\it{mass}} order parameter, suggesting it is directly related to the spectrum of massive excitations for $\beta  < \beta_c$. We note that the phase transition separates a massive phase, with $\Delta = 0$, from a line of critical points. This is the hallmark of the BKT phase transition, realized by the SG model. Therefore, we put forward the conjecture that the truncation prescription, implemented on the Langevin equations, is able to preserve the SG thermodynamical phases.
\vspace{-0.3cm}

\begin{figure}[tbph]
\includegraphics[width=9.5cm, height=8.5cm]{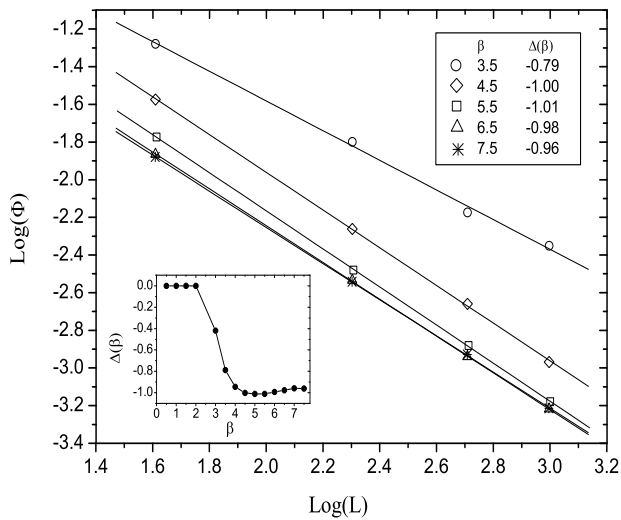}
\caption{Finite size scaling results obtained from the analysis of lattices with sizes $5 \times 5$, $10 \times 10$, $15 \times 15$ and $20 \times 20$. The inset shows the scaling dimension of the order parameter as a function of $\beta$.}
\label{fi43}
\end{figure}

An important difference to the SG physics calls attention, revealed when one examines the results of simulations for the truncated CDSG model: while the scaling dimension of the mass order parameter depends quadratically on $\beta$ in the massless phase of the SG model, in the truncated context, it is approximately constant. That is not a surprise, however. Once we have deformed the original model, we cannot attribute the same scaling properties to formally identical operators. Also, non-universal parameters, like $\beta_c$ are not supposed to be the same in the alternative model. Actually, we have
found $\beta_c \simeq 2.5$, which is smaller than the value predicted for the SG model.

\section{Conclusions}

The alternative, but equivalent, chirally decomposed version of the SG model yields an interesting stage for the test of numerical methods devoted to the analysis of systems governed by complex-valued actions. We have been able to show how a slight modification of the CDSG model at the level of the complex Langevin equations provides a stable numerical framework, allowing us to identify expected properties of the SG model, as the BKT transition. The essential idea of the truncation prescription, performed on the ``bad behaved" hyperbolic terms contained in the complex Langevin equations, is to reduce the degree of divergence of the unstable manifolds defined in the phase space, while preserving the thermodynamical phases of the original model. 

Since the CDSG model is similar to several relevant models found in condensed matter and high energy physics, it is likely that the rather elementary method of numerical stabilization addressed in this paper will be of utility in these contexts. Low-dimensional condensed matter bosonized models and the case of finite fermion density QCD are the ideal instances where an analogous line of numerical investigation could be applied.

Furthermore, alongside with the numerical approach, a dynamical renormalization group analysis of the truncated stochastic differential equations is in order, to put the results established here on a firmer ground. In particular, it is of fundamental importance to rederive the BKT renormalization group flow directly from the Langevin equations (\ref{lgeqsc}), a task which is under current investigation.

\acknowledgements
This work has been partially supported by CAPES and FAPERJ. M.M. would like to thank CNPq for a PROFIX fellowship.

\end{document}